\documentclass[aps,amsmath,amsfonts,floatfix,onecolumn,superscriptaddress]{revtex4-2}
\usepackage{graphicx}
\usepackage[colorlinks,linkcolor=blue,citecolor=blue,urlcolor=blue]{hyperref}
\usepackage{natbib}

\usepackage{ulem}

\begin{document}

\title{Emergence of coupling induced transparency by tuning purely dissipative couplings}
\author{Kuldeep Kumar Shrivastava$^*$, Moulik Deviprasad Ketkar$^*$, Biswanath Bhoi, Rajeev Singh\\ \normalsize Department of Physics, Indian Institute of Technology (Banaras Hindu University), Varanasi - 221005, Bharat (India) \\ \normalsize e-mail: kuldeep224@gmail.com, rajeevs.phy@iitbhu.ac.in\\ \normalsize $^{*}$ These authors contributed equally to this work.}

\begin{abstract}
Controlled transitions between coupling induced transparency (CIT) and coupling induced absorption (CIA) are effects of both fundamental importance as well as potential applications in various devices. We have explored these peculiar phenomena in multi-mode coupled hybrid quantum systems by considering a tunable mode (TM) and several static modes (SMs). The individual SMs and TM are designed such that they show CIA, but upon coupling different SMs we observe a transition from CIA to CIT. 
The observation is attributed to a subtle interplay of various couplings between TM and SMs, as well as among the SMs.
Quite remarkably we are able to achieve CIT using only purely dissipative couplings, whereas it is well known that CIT appears with coherent coupling.
We have developed a robust quantum theory based formalism which is able to capture the transition between CIA to CIT and have the capability to explain the inter-transition (CIT to CIA) as well as intra-transitions (CIA to CIA, CIT to CIT etc.) in a multimode hybrid quantum system all with just linear approach. 
A general model is developed for hybrid quantum systems having N modes.
We have then explicitly described two sets of hybrid systems, the first set is of three modes, 1TM coupled with 2SMs, and the second set is of four modes, 1TM coupled with 3SMs. Later we have generalised it for hybrid quantum systems having N number of modes. The results provide a pathway for designing hybrid systems that can control the group velocity of light, offering potential applications in the fields of optical switching and quantum information technology. Our finding and formulation that in a single hybrid quantum system we can achieve controllable inter-transitions and intra-transitions of CIT/ CIA may open a tool and guidance for its application in quantum technology and quantum materials as the TMs/SMs may be well extended to other real/ quasi-particles also.
\end{abstract}
\flushbottom
\maketitle
\section{Introduction}
 Light matter interactions, their interplay and control are central to physics and engineering due to their significant potential applications in quantum optics, telecommunications, quantum information technology, cavity quantum electrodynamics, quantum materials and the advancement of new physics \cite{wei2022towards,zhang2016advances, li2020hybrid,kimble2008quantum,xiang2013hybrid,maurya2024room}. One of the key phenomena of light-matter interaction is electromagnetically induced transparency (EIT), where a medium become transparent near the coupling center due to reduction in the resonance absorption. This effect is crucial in quantum computation and memory technologies \cite{lvovsky2009optical,wei2022towards,ma2017optical}, which signifies applications in light storage, optical delay, and optical stoppage. On the other hand, electromagnetically induced absorption (EIA) which results in enhanced absorption of transmission signals have pivotal role in the applications such as fast light, molecular detection, photodetectors and, photovoltaics.

Analogous of both these phenomena, EIT and EIA, in the linear regime are the coupling-induced transparency (CIT) and absorption (CIA) respectively. A special form of light-matter interaction originates from phase relationship between different modes of the system or different subsystems used to exchange energy in a coherent way. Thus the coherent coupling is present everywhere in the form of EIT/ CIT characterized by their transmission profile having level repulsion (LR) in dispersion and attraction in their linewidth profile \cite{lezama1999electromagnetically, ma2017optical, brazhnikov2005electromagnetically, lv2023broadband, sun2022circularly, fleischhauer2005electromagnetically, maurya2024room}.

Another form of light-matter interaction which is relatively less explored and is now drawing much attention arises due to dissipative coupling in the form of EIA\,/~CIA, arising when energy of a system is getting significantly dissipated at some frequencies.  
This phenomena manifests itself as level attraction (LA) in transmission profile of dispersion and repulsion in their linewidth. Dissipative coupling is the backbone for developing nonreciprocal devices for enhancing sensing techniques, optical isolators. 
It may also be used to quantify degradation of entanglement, coherent loss etc \cite{lezama1999electromagnetically,brazhnikov2005electromagnetically,lv2023broadband,sun2022circularly}. 
In order to have good control over light matter interaction, understanding electromagnetic behaviour at the microscopic level and controlling both coherent and dissipative coupling is one of the promising ways which is also required for advancing future quantum technologies and their applications.
In an earlier work \cite{shrivastava2024unveiling}, we have observed both CIT and CIA in a single planar device by just changing the geometry of resonators. Here we present a mechanism by which these phenomena can be interchanged by controlling the coupling between the resonators. We briefly summarize these two phenomena and explain their differences before discussing our mechanism \cite{shrivastava2024unveiling,bhoi2022coupling,harder2021coherent}.

\begin{figure}
        \centering
        \includegraphics[width=\linewidth]{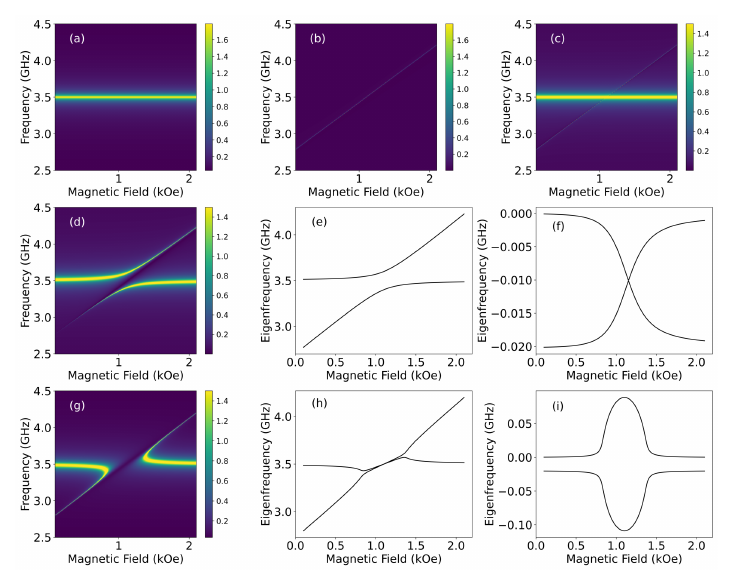}
        \caption{Transmission profile for (a) photonic modes, and (b) magnonic mode. (c) Transmission profile for combined photonic and magnonic modes without any coupling. Coherent coupling (d, e, f), (d) transmission profile showing repulsive dispersion in nature, (e) real part of eigenvalue showing level repulsion and (f) imaginary part of eigenvalue showing attraction in linewidth. Dissipative coupling (g, h, i), (g) transmission profile showing attractive dispersion in nature, (h) real part of eigenvalue showing level attraction and (i) imaginary part of eigenvalue showing repulsion in linewidth.}
        \label{F1C5}
    \end{figure}  

{\bf Coupling induced transparency (CIT): Level repulsion/ normal anti-crossing}
In coupling induced transparency, coherent coupling is characterized by its dispersion relation that shows level repulsion in the transmission profile of the hybridised modes around the coupling centre as shown in, Fig.1(d). The real and imaginary parts of the complex eigenvalues show contrasting behaviour. The real part exhibit level repulsion around the coupling centre, resulting in distinct upper and lower branches at the center ( Fig.1(e)), while the imaginary part, which represents the linewidth profile, shows crossing or attraction (Fig.1(f)). This kind of coupling is very ubiquitous in nature and plays an important role in applications involving transduction, allowing efficient energy exchange between two modes.

{\bf Coupling induced absorption (CIA): Level attraction/ opposite anti-crossing}
In coupling induced absorption, transmission profile of the phenomena near the coupling centre is characterised by dissipative nature of the coupling and shows attraction in the dispersion profile Fig.1(g). The real part shows level attraction around the coupling centre, Fig.1(h), i.e. the upper and lower branches of the eigenfrequencies will get merged at/ near the coupling centre. The imaginary part representing its linewidth profile on the other hand shows repulsion, Fig.1(i). Recently a distinct magnon–photon dissipative coupling was discovered and it has been quickly verified in a variety of setups with different cavity configurations \cite{harder2021coherent,bhoi2022coupling}. A distinct feature of a dissipatively coupled system is the level attraction (LA) of the hybridized modes as shown in Fig.1(g,h) and repulsion in their linewidth Fig.1(i), which is in strong contrast to the results induced by coherent coupling.

Another aspect of CIT and CIA that has caught some attention recently is the transition between them through coupling of other modes in multimode hybrid systems. In a previous study, Bhoi et al. reported a planar structure in which one magnonic mode (YIG) was coupled with three ISRR photonic modes \cite{bhoi2022coupling}. The ISRR photonic modes were concentric square rings in the ground plane and YIG lied on the centre of these concentric rings but on the front plane touching the microstrip line.  Individual square rings were of different dimensions having different resonance frequencies. When the individual rings were forming the hybrid system with YIG they showed level attraction at their respective resonance frequencies when YIG resonance frequency reached near ISRR resonance frequency by tuning the magnetic field. But when the three concentric ISRRs were combined with YIG together the behaviour of their dispersion profile at the three crossings were different, one showing level repulsion, another showing level absorption and one showing level attraction. A classical theory was used to explain the phenomena, but here we show that this phenomenon is possible even in the quantum domain by developing a full quantum theory for similar observations.

Another study \cite{hu2022auxiliary} described a cavity magnonic system of three modes in which the coupling between cavity and magnon modes were mediated by an SRR auxiliary mode. By controlling the damping of auxiliary mode the group has achieved both normal (CIT/LR) and opposite (CIA/LA) anti-crossing. In yet another experiment \cite{rao2020interactions} with a hybrid system composed of a single YIG sphere and two orthogonally crossed coplanar waveguide they observed both normal (CIT/LR) and opposite (CIA/LA) anti-crossing. They have also analysed how destructive interference between magnon-dipole and magnon-quadruple determines the interactions and have discussed perfect absorption at the opposite anti-crossing in the $S_{21}$ spectrum.

Extending the approach of our previous work \cite{shrivastava2024unveiling}, we have applied the quantum theory to the CIT and CIA for hybrid systems having multiple modes and explored multiple phenomena of CIA and CIT in a single device and also achieved transition between them.
In our study we have developed a quantum theoretical framework for the hybrid quantum systems having multiple modes. 
We are using two different modes one having tunable property (TM) and other modes are static (SM) in the nature. 
We have developed a general model for multiple modes having different hybrid quantum systems of N modes.
We are considering two cases the first case having three modes, 1TM and 2SMs. When individual SM is getting coupled with TM, LA is happening but when both SMs are getting coupled with TMs we observe transition from LA to LR. The second case has four modes, 1TM and 3SMs. Again, when individual SM is getting coupled with TM, LA is happening but when all SMs are getting coupled with TM the transition from LA to LR is happening. When the transition from LA to LR is happening in the midway of the transition passing from level absorption is also observed. Here for the two simple cases, we achieved controllable transition from LA to LR in the multimode hybrid quantum system by tuning the coupling parameters. Different nature of transitions e.g., from LR to LA, LR to LR, LA to LA, LA or LR to level absorptions and vice versa may also be achieved with different sets of parameters. Many groups have already observed some of these phenomenon. Our findings unveil the complex dynamics of the interaction between different modes of hybrid quantum systems which may be expected to advance the applications for future quantum technologies.

\section{General model}
 
Different coupled mode of a hybrid system connected to a channel additionally affect each other through the channel, which can be in the form of bath\,/ cavity\,/ microstripline etc., to inject travelling photons in the system and is  also attached to input and output ports. We may write a general Hamiltonian for such a system as \cite{manasi2018light,tiwari2024modified,rao2020interactions,harder2021coherent,walls2008quantum,scully1997quantum,shrivastava2024unveiling,hu2022auxiliary}

\begin{align} \label{E1C5}
    H / \hbar = \sum_{l=1}^{N}(\omega_l -i\alpha_l)\hat X_l ^\dag \hat X_l + \sum_{1\leq l < m \leq N} \Delta_{lm}(\hat X_l +\hat X_l^\dag)(\hat X_m +\hat X_m^\dag)  +\int\omega_k \hat p_k^\dag \hat p_k dk +\int\left[\sum\lambda_l (\hat X_l +\hat X_l^\dag)(\hat p_k+\hat p_k^{\dag})\right] dk.
 \end{align} 
After taking rotating wave approximation (RWA) we may write Eq.~\ref{E1C5} as

\begin{align} \label{E2C5}
    H / \hbar = \sum_{l=1}^{N}(\omega_l -i\alpha_l)\hat{X_l}^{\dag}\hat{X_l} +\sum_{1\leq l < m \leq N}\Delta_{lm}(\hat{X_l} \hat X_m ^\dag + \hat X_m \hat X_l ^{\dag})  +\int\omega_k \hat p_k^{\dag} \hat p_k dk+\int\left[\sum\lambda_l (\hat{X_l} \hat p_k^{\dag} +\hat{X_l}^{\dag}\hat p_k)\right]dk.
 \end{align} 

\noindent Here $l$ is the index of the modes in the system which is varying from $1$ to $N$, $\hat{X}^\dag_l(\hat{X_l})$ is the creation (annihilation) operators of the mode $l$. $\omega_l$ denote the resonance frequencies while $\alpha_l$ denote the intrinsic damping rate of the uncoupled mode $l$, the coupling parameter between different modes $l$ and $m$ are denoted by $\Delta_{lm}=J_{lm}+i\Gamma_{lm} $  where $J_{lm}$ and $\Gamma_{lm} $ are real parameters that characterize the strength of coherent and dissipative interactions between the modes $l$ and $m$. 

The third term of the Hamiltonian represents the feeding channel 
(cavity\,/~bath\,/~microstripline etc. as the case may be) connected to the input and output ports. In our formulation of Hamiltonian, we have modelled the feeding channel through which traveling photons mediate the interaction, integrating over a real domain from $-\infty $ to $+\infty $. Bosonic creation (annihilation) operator of the traveling photon is denoted by ${\hat p_k}^\dag (\hat p_k)$ which obeys
$[\hat p_k, \hat p_{k'}^\dag] = \delta(k-k')$.
$\omega_k$ denotes the frequency of travelling photon where $k$ represents the wave vector. The last term represents the interaction between each mode and traveling photons, the interaction strength between them is denoted by $\lambda_l$, modelled linearly in ${\hat p_k}^\dag (\hat p_k)$.

Following standard input output formalism, the time forwarded Heisenberg-Langevin equations \cite{shrivastava2024unveiling,rao2020interactions,harder2021coherent} of the $l$-th mode of the coupled system reads,

\begin{align} \label{E3C5 }
\dot{\hat{X_l}} (t)  = -i\Tilde{\omega_l}  \hat{X_l} (t)- \beta_l \hat{X_l} (t) - i \sqrt{\beta_l}\hat P_{in} (t)  -\sum_{\substack{m=1\\ m \neq l}}^N \sqrt{\beta_l\beta_m } \hat X_m (t) - i\sum_{\substack{m=1\\ m \neq l}}^N \Delta_{lm} \hat X_m (t)
\end{align} 
where $\Tilde{\omega_l} = \omega_l - i\alpha_l$, $\beta_l = 2 \pi\lambda_l^2 $ and $\beta_m = 2 \pi\lambda_m^2 $ represent the extrinsic damping rates of the mode $l$ and $m$ respectively. Here $\hat P_{in}$ ($\hat P_{out}$) is defined as the input (output) field operator at the input (output) port. We observe that the external damping rates ($\beta$'s) not only cause additional damping on top of the internal dissipation of respective components, but also cause the coupling between the components to become more dissipative. After applying the Fourier transform the frequency domain equation may be written as \cite{shrivastava2024unveiling,rao2020interactions,harder2021coherent}
\begin{align} \label{E4C5 }
 i(\omega - &\Tilde{\omega_l} ) \hat{X_l} (\omega)- \beta_l \hat{X_l} (\omega) - i \sqrt{\beta_l}\hat P_{in} (\omega)-\sum_{\substack{m=1\\ m \neq l}}^N \sqrt{\beta_l\beta_m } \hat X_m (\omega) - i\sum_{\substack{m=1\\ m \neq l}}^N \Delta_{lm} \hat X_m (\omega) = 0
\end{align} 
Similarly, the time retarted Heisenberg-Langevin equation \cite{shrivastava2024unveiling,rao2020interactions,harder2021coherent} of the coupled system in the time and frequency domain reads
\begin{align} \label{E5C5 }
\dot{\hat{X_l}} (t)  &= -i\Tilde{\omega_l}  \hat{X_l} (t)+ \beta_l \hat{X_l} (t) - i \sqrt{\beta_l}\hat P_{out} (t) +\sum_{\substack{m=1\\ m \neq l}}^N\sqrt{\beta_l\beta_m } \hat X_m (t) - i\sum_{\substack{m=1\\ m \neq l}}^N \Delta_{lm} \hat X_m (t)
\end{align} 

\begin{align} \label{E6C5 }
 i(\omega - &\Tilde{\omega_l} ) \hat{X_l} (\omega)+ \beta_l \hat{X_l} (\omega) - i \sqrt{\beta_l}\hat P_{out} (\omega) +\sum_{\substack{m=1\\ m \neq l}}^N\sqrt{\beta_l\beta_m } \hat X_m (\omega) - i\sum_{\substack{m=1\\ m \neq l}}^N \Delta_{lm} \hat X_m (\omega) = 0
\end{align}
We need to figure out the algebraic relation between the input and output port variables in order to derive the transmission results. From Eqs.~\ref{E4C5 }~and~\ref{E6C5 } we can get the input and output field relation as
\begin{align} \label{E7C5}
\hat P_{out} (\omega)=\hat P_{in} (\omega) - 2i \sum_{i=1}^{N} \sqrt{\beta_l} \hat{X_l} (\omega)
\end{align}
To study these effects we measure the transmission coefficient $S_{21}$ between input and output port of the hybrid system, which is defined as
\begin{align} \label{E8C5}
S_{21} = \frac{\hat p_{out}}{\hat p_{in}} - 1
\end{align}
Solving Eqs.~\ref{E4C5 }, ~\ref{E7C5}, and~\ref{E8C5} 
numerically gives the transmission profile. $S_{21}$ may also be written in matrix form as follows
\begin{align} \label{E9C5}
S_{21} = \mathcal{B}^T\mathcal{M}^{-1}\mathcal{B}
\end{align} 
where 
\begin{center}
 $\mathcal{B} =\sqrt{2}\begin{bmatrix}
    \sqrt{\beta_1}\\
    \sqrt{\beta_2}\\
    .\\
    .\\
    .\\
    \sqrt{\beta_N}
     \end{bmatrix}_{N \times 1}$  ,   
\end{center} ~\\  
\begin{center}
     $\mathcal{M} = i\begin{bmatrix}
    &\omega-\tilde{\tilde{\omega}}_1 
    &-\Delta_{12}+i\sqrt{\beta_1 \beta_2} &.&.&.
    &-\Delta_{1N}+i\sqrt{\beta_1 \beta_N}\\
    &-\Delta_{12}+i\sqrt{\beta_1 \beta_2} 
    &\omega-\tilde{\tilde{\omega}}_2  &.&.&.
    &-\Delta_{2N}+i\sqrt{\beta_2 \beta_N}\\
    &.&.&.&.&.&.\\
    &.&.&.&.&.&.\\
    &.&.&.&.&.&.\\
    &-\Delta_{1N}+i\sqrt{\beta_1 \beta_N}
    &-\Delta_{2N}+i\sqrt{\beta_2 \beta_N}
    &.&.&.
    &\omega-\tilde{\tilde{\omega}}_N
     \end{bmatrix}_{N \times N}$ 
\end{center}
     also
     $\Delta_{lm}=\Delta_{ml}$ , and
     $\tilde{\tilde{\omega_l}}=\tilde{\omega_l} - i\beta_l=\omega_l - i(\alpha_l +\beta_l)$.
     
To keep the discussion more concrete experimentally, we are assuming TM to be a magnon mode, magnon being collective spin excitation of ferromagnetic/antiferromagnetic material, resonance frequency of which can be tuned by applying external magnetic field. We may choose some magnetic thin film (e.g. yttrium iron garnet (YIG) ) for excitation of the magnon mode, representative transmission profile is shown in Fig.~\ref{F1C5}(b). Similarly for  SM modes we may opt for photonic resonators (e.g. split ring resonator (SRR) / inverted split ring resonator (ISRR)) resonance frequencies of which will remain constant with the applied magnetic field. Transmission profile is shown in the Fig.~\ref{F1C5}(a) for one photonic mode. Fig.~\ref{F1C5}(c) shows the case when there is no coupling between the photonic and magnonic modes. 

\section{Three mode coupled hybrid quantum systems}

\begin{figure}[h!]
    \centering
    \includegraphics[width = \linewidth]{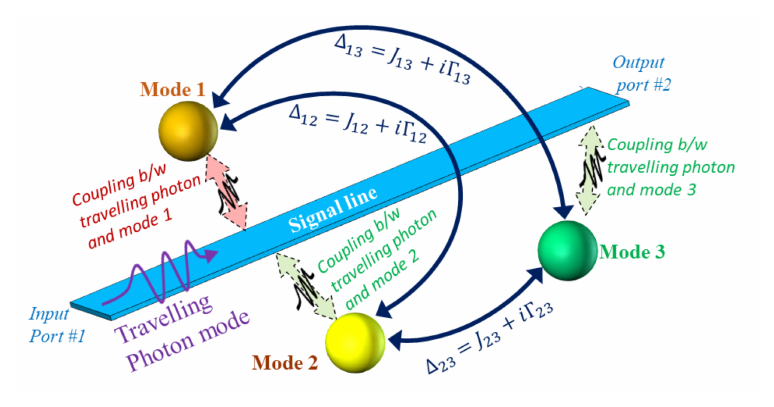}
    \caption[Cartoon showing an example of the hybrid system having three modes]
    {Cartoon showing a simple planar hybrid system having three modes, out of which mode~1 ($M$) is tunable, and mode~2 ($P_1$) and mode~3 ($P_2$) are static modes. For tuning of $M$ we have applied an external inplane magnetic field perpendicular to the feedline (MSL).}
    \label{F2C5}
\end{figure}

The simplest case where a transition between CIT and CIA is possible without changing the nature of any coupling, is that of three modes. For simplicity we consider one of the modes to be tunable while the other two as static. Schematic of such an experiment is shown in Fig.~\ref{F2C5}, where the tunable mode we are taking as magnon (YIG) that we are denoting by M, whose resonance frequency can be tuned by applying an external magnetic field. Static modes are made out of ISSRs, whose resonance frequencies will remain unaffected by the applied external magnetic field. The two static modes are two different ISRRs of different dimensions, having different resonance frequencies. The feedline in this planar structure is a microstripline (MSL) common to both type of modes, which in turn is also attached to the input and output ports of a vector network analyzer (VNA), from where microwave photons are getting injected into the system that are initiating and creating all of the dynamics.

\subsection{Model and analysis}

We have modelled the three mode coupled photon-magnon hybrid quantum system  using the Hamiltonian of Eq.~\ref{E2C5} by restricting it to 1 magnon ($M$) and 2 photon modes $P_1$  and  $P_2$ \cite{hu2022auxiliary,tiwari2024modified, rao2020interactions,harder2021coherent, walls2008quantum,scully1997quantum,shrivastava2024unveiling,manasi2018light}
\begin{align} \label{E10C5}
    H_{(M, P_1, P_2)} / \hbar &=\tilde{\omega}_M \hat X_M^\dag \hat X_M + \tilde{\omega}_{P_1} \hat X_{P_1}^\dag \hat X_{P_1} + \tilde{\omega}_{P_2}\hat X_{P_2}^\dag \hat X_{P_2} + \int\omega_k \hat p_k^\dag \hat p_k dk  \nonumber \\
    & +\Delta_{MP_1}(\hat X_M \hat X_{P_1}^\dag + \hat X_{P_1} \hat X_M^\dag)+
    \Delta_{MP_2}(\hat X_M \hat X_{P_2}^\dag + \hat X_{P_2} \hat X_M^\dag)+
    \Delta_{P_1 P_2}(\hat X_{P_1} \hat X_{P_2}^\dag + \hat X_{P_2} \hat X_{P_1}^\dag)\nonumber \\
    & + \int\left[\lambda_M (\hat X_M \hat p_k^\dag +\hat X_M^\dag \hat p_k)+\lambda_{P_1} (\hat X_{P_1} \hat p_k^\dag +\hat X_{P_1}^\dag\hat p_k)+\lambda_{P_2} (\hat X_{P_2} \hat p_k^\dag +\hat X_{P_2}^\dag \hat p_k)\right]dk , 
 \end{align} 
 where $\tilde{\omega}_M = \omega_M -i\alpha_M$, $\tilde{\omega}_{P_1} = \omega_{P_1} -i\alpha_{P_1}$ and $\tilde{\omega}_{P_2} = \omega_{P_2} -i\alpha_{P_2}$ are the complex frequencies of the magnon $M$ and photon modes $P_1$ and $P_2$ respectively. Here $\alpha_M $ , $\alpha_{P_1} $ and $\alpha_{P_2} $ are intrinsic damping parameters for the $M$, $P_1$ and $P_2$ modes respectively. It is a well established now that LA, or LR is determined by the relative and combined strengths and phase of the oscillating magnetic fields generated from ISRR's split gap, magnon and the travelling waves of MSL. Also for coherent coupling (LR) the complex coupling constant ($\Delta= J+i\Gamma$) has a dominating real part ($J$) while for dissipative coupling (LA) it has a dominating imaginary part ($\Gamma$). 
 
 The Heisenberg-Langevin equations for this system can be compactly written in a matrix form as
 \begin{align} \label{E11C5}
     \frac{d}{dt} \begin{bmatrix} \hat X_M\\ \hat X_{P_1}\\ \hat X_{P_2}
     \end{bmatrix}
     = -i\,H_{coupling}
      \begin{bmatrix} \hat X_M\\ \hat X_{P_1}\\ \hat X_{P_2}
     \end{bmatrix}
     -i\begin{bmatrix} \sqrt{\beta_M}\\\sqrt{\beta_{P_1}}\\\sqrt{\beta_{P_2}}
     \end{bmatrix} \hat P_{in} (t)
 \end{align} 
where, 
 \begin{center}
 $ H_{coupling}= \begin{bmatrix}
    &\tilde{\tilde{\omega}}_M 
    &\Delta_{MP_1}-i\sqrt{\beta_M \beta_{P_1}} &\Delta_{MP_2}-i\sqrt{\beta_M \beta_{P_2}}\\
    &\Delta_{MP_1}-i\sqrt{\beta_M \beta_{P_1}} 
    &\tilde{\tilde{\omega}}_{P_1} &\Delta_{P_1 P_2}-i\sqrt{\beta_{P_1} \beta_{P_2}}\\
    &\Delta_{MP_2}-i\sqrt{\beta_M \beta_{P_2}}
    &\Delta_{P_1 P_2}-i\sqrt{\beta_{P_1} \beta_{P_2}}
    &\tilde{\tilde{\omega}}_{P_2}
     \end{bmatrix}$    
 \end{center} 
     
 \noindent where, $\beta_M $, $\beta_{P_1}$, $\beta_{P_2}$ are external damping rates for the $M$, $P_1$ and $P_2$ modes respectively and $\Delta_{lm}=\Delta_{ml}$ and
 $\tilde{\tilde{\omega_l}}=\tilde{\omega_l} - i\beta_l=\omega_l - i(\alpha_l +\beta_l)$. Explicitly the complex frequencies of the modes after taking into account the internal damping and coupling induced (external) damping may be written as  $\tilde{\tilde{\omega}}_M=\omega_M - i(\alpha_M +\beta_M)$ , $\tilde{\tilde{\omega}}_{P_1}=\omega_{P_1} - i(\alpha_{P_1} +\beta_{P_1})$ , $\tilde{\tilde{\omega}}_{P_2}=\omega_{P_2} - i(\alpha_{P_2} +\beta_{P_2})$ for magnon mode ($M$), photon modes $P_1$ and $P_2$ respectively. Whenever the  direct interaction between magnon and the two photon modes happens, the resonance frequencies of the hybridised modes are splitted into higher and lower branches for LR and get merged around coupling center for LA, that can be analytically solved by using coupling matrix ($H_{coupling}$) of the Eq.~\ref{E11C5} for real parts of its eigenvalues. 

 In this case using  Eq.~\ref{E7C5}, for three modes we can write $P_{out}$ and $P_{in}$ relation in frequency domain as \cite{rao2020interactions,hu2022auxiliary,shrivastava2024unveiling}
\begin{align} \label{E12C5 }
\hat P_{out}(\omega)-\hat P_{in}(\omega)= - 2i \left[\sqrt{\beta_M} \hat{X}_M (\omega) + \sqrt{\beta_{P_1}} \hat X_{P_l} (\omega) + \sqrt{\beta_{P_2}} \hat X_{P_2} (\omega)\right] 
\end{align}

For calculating the dispersion spectra of different types of cooperative effect taking into account all the interactions of magnon mode, photon modes and travelling photon modes of the combined hybrid system first we need to calculate $S_{21}$ (on $\frac{\omega}{2\pi} - H_{dc}$ plane). Following Eq.~\ref{E8C5} this can be written for 1 magnon and 2 photon modes as
\begin{align} \label{E13C5 }
S_{21}=\frac{\hat P_{out}}{\hat P_{in}}-1= - \frac{2i}{\hat P_{in}} \left[\sqrt{\beta_M} \hat{X}_M + \sqrt{\beta_{P_1}} \hat X_{P_l} + \sqrt{\beta_{P_2}} \hat X_{P_2}\right] 
\end{align}
The transmission profile in matrix form for 3 modes is given by Eq.~\ref{E9C5} with
\begin{align} \label{E14C5}
S_{21} = \mathcal{B}^T_{1\times 3}\mathcal{M}_{3\times 3}^{-1}\mathcal{B}_{3\times 1}
\end{align} 
where

\begin{center}$\mathcal{B}=\sqrt{2}\begin{bmatrix}
    \sqrt{\beta_M}\\
    \sqrt{\beta_{P_1}}\\
    \sqrt{\beta_{P_2}}
     \end{bmatrix}$ ,
\end{center}

\begin{center}
$\mathcal{M}=i\begin{bmatrix}
    &\omega-\tilde{\tilde{\omega}}_M 
    &-\Delta_{MP_1}+i\sqrt{\beta_M \beta_{P_1}}
    &-\Delta_{MP_2}+i\sqrt{\beta_M \beta_{P_2}}\\
    &-\Delta_{MP_1}+i\sqrt{\beta_M \beta_{P_1}} 
    &\omega-\tilde{\tilde{\omega}}_{P_1}
    &-\Delta_{P_1 P_2}+i\sqrt{\beta_{P_1} \beta_{P_2}}\\
    &-\Delta_{MP_2}+i\sqrt{\beta_M \beta_{P_2}}
    &-\Delta_{P_1 P_2}+i\sqrt{\beta_{P_1} \beta_{P_2}}
    &\omega-\tilde{\tilde{\omega}}_{P_2}
     \end{bmatrix}$.    
\end{center}

\subsection{Results and discussion}

\begin{figure}
  \centering
  \includegraphics[width = 0.88\linewidth]{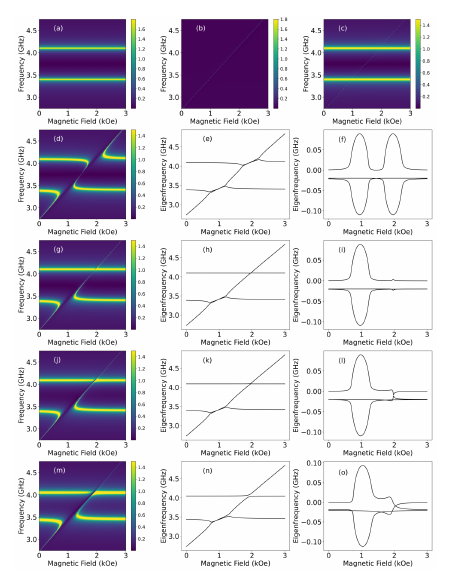}
    \caption {Hybrid quantum system having 1 TM coupled individually to 2 SMs. (a) Transmission profile for two static modes without magnon mode, (b) transmission profile for magnon mode without static modes, (c) combined transmission profile for two static modes and magnonic mode without any coupling.  
    Fig(d) Transmission profile, (e) real, and (f) imaginary part of complex eigenfrequencies, with no coupling between individual SMs.
   (g) Transmission profile; (h) real, and (i) imaginary part of complex eigenfrequencies, when coupling parameter between SMs is $0.01i$.
   (j) Transmission profile; (k) real, and (l) imaginary part of complex eigenfrequencies, when coupling parameter between SMs is $0.1i$.
   (m) Transmission profile; (n) real, and (o) imaginary part of complex eigenfrequencies, when coupling parameter between SMs is $0.2i$. \\
   The change of parameters and eigenvalue properties are summarized in Table~\ref{T1C5}.} \label{F3C5}
\end{figure}

The transmission properties of the system are studied using vector network analyzer which sends microwave photons of frequency $f_{AC} = \omega / 2\pi$ through the feeding line MSL for different strengths of  applied static magnetic field $H_{dc}$.
Fig.~\ref{F3C5}(a) shows $S_{21}$ power spectra of the two ISRRs without YIG film, where only two transmission lines horizontal to the applied $H_{dc}$ are visible in the transmission profile on the $(H_{dc} -f_{AC})$ plane which is  unaffected by the tuned magnetic field.
The horizontal lines are at resonance frequencies $\omega_{P_1}$= 3.4, $\omega_{P_2}$= 4.1 GHz with intrinsic damping rates $\alpha_{P_1}$=0.002, and $\alpha_{P_1}$=0.002 respectively \cite{bhoi2022coupling,bernier2018level,hu2022auxiliary,rao2020interactions,harder2021coherent}. 
Fig.~\ref{F3C5}(b) shows $S_{21}$ power spectra of the YIG film without the ISRRs, showing magnon excitations in the YIG films termed as ferromagnetic resonance (FMR) mode having intrinsic damping rate $\alpha_{M}$=0.00002 \cite{bhoi2022coupling,bernier2018level,hu2022auxiliary,rao2020interactions,harder2021coherent}, visible as a slant transmission line which is getting tuned by the applied $H_{dc}$ ranging from $0.0$ to $3.0$ $kOe$.  Resonance frequency of this FMR mode can be linearly modelled as $\omega_{M} = 0.714\times H_{dc}+2.714$ which is varying from 2.714 to 4.856 GHz. Fig.~\ref{F3C5}(c) shows $S_{21}$ power spectra of the two ISRRs and YIG film in absence of any coupling, which is approximately the addition of Fig.~\ref{F3C5}(a) and~(b), and the transmission line of magnon simply crosses the transmission lines of photons.

When YIG gets excited and its FMR mode interacts with $P_1$ and $P_2$ modes with some strength of coupling parameter, its effects will be reflected in the transmission profile. When only $M$ and $P_1$ ($P_2$) interact with the coupling parameter $\Gamma_{MP_1} $ = 0.1$i$ ($\Gamma_{MP_2} $ = 0.1$i$) there will be LA 
near the resonance frequency of the coupling center of the coupled pair for $MP_1$ at 3.4 GHz ($MP_2$ at 4.1 GHz) as shown in Fig.~\ref{F3C5}(d). Fig.~\ref{F3C5}(e) shows the real part of the complex eigenfrequencies of the state of the system. This shows two clear and distinct LAs, one each for the $MP_1$ and $MP_2$ where higher and lower eigenfrequency branches are getting merged and then separated. 
In these two cases of LAs the lower and upper branches of the hybridised modes are getting attracted and merging to a common eigenfrequency near the coupling center with comparatively strong microwave absorption, resulting in an anti-crossing called CIA. 

In the hybrid system multiple interaction paths between the travelling waves of MSL, magnon mode  and two photon modes are giving rise to the phenomenon of LA and LR. Fig.~\ref{F3C5}(d,e) shows the decoupled case between the photonic modes (i.e.~$\Delta_{P_1 P_2} = 0$). The travelling microwave photons are reaching the different photonic modes and the magnonic mode is getting directly coupled with the two photonic modes, resulting in further interaction with the travelling photon modes of MSL. In this way the hybrid coupling modification with the travelling waves of photon modes will determine the type of interactions, LA or LR etc. 

For the first part of coupled hybrid system when we are considering interactions between $M$ and $P_1$, their dispersion [Fig.~\ref{F3C5}(d)] and real part of their eigenvalue [Fig.~\ref{F3C5}(e)] are in good agreement for the LA at 3.4 GHz with internal and external dissipation 0.00002 and 0.00018 (0.002 and 0.018) \cite{bhoi2022coupling, bernier2018level, hu2022auxiliary, rao2020interactions, harder2021coherent} for magnon (photon 1) mode respectively. The internal and external dissipations for Fig.~\ref{F3C5}(d-o) are going to be constant throughout the discussions.

In Fig.~\ref{F3C5}(f) we have drawn the imaginary part of complex eigenvalues solution of $H_{coupling}$ for $MP_1$ and $MP_2$. 
For $MP_1$ first part of Fig.~\ref{F3C5}(e, h, k, n) are real part of their complex eigenvalues that also correspond to the first part of transmission profile Fig.~\ref{F3C5}(d, g, j, m) respectively and show LA. Repulsive nature in linewidth for Fig.~\ref{F3C5}(f, i, l, o) confirms dissipative coupling (CIA) having the coupling constant $ \Delta_{MP_1} = 0.1i$ for the first part of the Fig.~\ref{F3C5}(d-o).
The coupling constant $ \Delta_{MP_1} = 0.1j$ corresponds to the Fig.~\ref{F3C5}(d, e, f), where Fig.~\ref{F3C5}(d, e) shows lower\,/~higher frequency branches of the corresponding hybrid modes. These two modes are merged near coupling center because of CIT phenomenon by virtue of imaginary coupling parameter. 
Fig.~\ref{F3C5}(d, g, j, m) representing transmission profiles and their corresponding real  and imaginary parts of complex eigenvalue are shown in Fig.~\ref{F3C5}(f, i, l, o) respectively. 

We now consider interactions between $M$ and $P_2$ in second part of the coupled hybrid system taking coupling constant $ \Delta_{MP_2} = 0.1i$ corresponding to Fig.~\ref{F3C5}(d,e) where lower/higher branches of the frequency correspond to the hybrid modes, the two modes are merged near coupling center because of CIA effects by having imaginary coupling parameter between $M$ and $P_2$, and with no coupling between $P_1$ and $P_2$. 
The dispersion profiles shown in Fig.~\ref{F3C5}(d, g, j, m) and real part of their eigenvalues Fig.~\ref{F3C5}(e, h, k, n) agrees well for both at 3.7 GHz with internal and external dissipation 0.00002 and 0.00018 (0.002 and 0.018) for magnon (photon 2) mode \cite{bhoi2022coupling,bernier2018level,hu2022auxiliary,rao2020interactions,harder2021coherent} respectively. The internal and external dissipations for $MP_2$ are going to be constant throughout the discussion. 

\begin{table}   [h!]
\begin{center}
\begin{tabular}{ |c|c|c|c|c| } 
 \hline
 Subplots & $\Delta_{MP_2}$ & $\Delta_{P_1P_2}$ & \multicolumn{2}{c|}{Eigenvalues near $MP_2$ crossing}\\
 \cline{4-5}
 & & & Real part & imaginary part\\
 \hline
 d-f & 0.1i & 0i & Attraction & Repulsion\\ 
 g-i & 0.01i & 0.01i & Intermediate & Repulsion\\ 
 j-l & 0.01i & 0.1i & Intermediate & intermediate\\
 m-o & 0.02i & 0.2i & Repulsion & Attraction\\
 \hline
\end{tabular}
\end{center}
\caption{Set of parameters that are changing and the eigenvalues which are changing qualitatively in Fig.~\ref{F3C5}} 
    \label{T1C5}
\end{table}

The coupling constant  between $M$ and $P_2$ for the third row Fig.~\ref{F3C5}(g, h, i) is $\Delta_{MP_2} = 0.01i$  and the inter-photonic coupling constant is $\Delta_{P_1 P_2} = 0.01i$ between $P_1$ and $P_2$. Fig.~\ref{F3C5}\,(g,h) shows lower\,/ higher branches of the eigenfrequencies corresponding to the hybrid modes. These coupled modes are crossing near coupling center implying CIA phenomenon but now having a less imaginary coupling parameter. In Fig.~\ref{F3C5}(i) we have drawn imaginary part of the complex eigenvalues of the coupling matrix, for $MP_2$  showing that the repulsion in their linewidth at the coupling centre have reduced considerably compared to the case Fig.~\ref{F3C5}(f).
The coupling constant  between $M$ and $P_2$ for the fourth row Fig.~\ref{F3C5}(j, k, l) is $\Delta_{MP_2} = 0.01i$ with inter-photonic coupling constant $\Delta_{P_1 P_2} = 0.1i$ between $P_1$ and $P_2$. Fig.~\ref{F3C5}(j,k) shows lower\,/~higher frequency branches of the corresponding hybrid modes, these two modes are still crossing near coupling center and still confirming CIA phenomenon by virtue of a reduced imaginary coupling strength despite of increasing inter-photonic coupling constant. In Fig.~\ref{F3C5}(l) we have drawn imaginary part of the eigenvalues of $H_{coupling}$ for $MP_2$, in their linewidth the repulsion diminishes and it is just to start attraction/ crossing but have not started yet, this behaviour still confirms CIA. 
The coupling constant  between $M$ and $P_2$ for the fifth row Fig.~\ref{F3C5}(m, n, o) is $\Delta_{MP_2} = 0.02i$ and inter-photonic coupling constant is $\Delta_{P_1 P_2} = 0.2i$ between $P_1$ and $P_2$. Fig.~\ref{F3C5}(m,n) shows lower/higher frequency branches of the coupled hybrid modes with no attraction in the levels but they remain separated near coupling center that clearly confirms CIT phenomenon that is combined effect of $\Delta_{MP_2}$ and $\Delta_{P_1 P_2}$, which is now twice the value from the previous row Fig.~\ref{F3C5}(j, k, l). In Fig.~\ref{F3C5}(l) we have drawn the imaginary part of  eigenfrequencies of $H_{coupling}$, showing clear attraction in their linewidth as they are crossing each other. 

We want to emphasize here that all the coupling parameters are pure imaginary and individually will result in LA, but the combined effect for suitable choice of parameters is LR with only dissipative couplings. The observation of CIA(LA) and CIT(LR) may be described as a balance between the multiple interaction possibilities between the magnon mode, photon modes and travelling photons. In the region where magnon mode and first photon mode are getting coupled around 3.4 GHz, their respective internal dissipation and external dissipation are constant, these parameters are not going to play much role in the transition of the behaviour. Also coupling parameter between $M$ and $P_1$ i.e. $\Delta_{MP_1}$ is not changing from second to fifth row so its behaviour in transmission profile and in real or imaginary parts of eigenvalues is also not getting affected, although coupling parameters between $M$ and $P_2$ i.e. $\Delta_{MP_2}$ and $P_1P_2$ i.e. $\Delta_{P_1P_2}$ are changing but the effective contribution of $\Delta_{MP_2}$ and $\Delta_{P_1P_2}$ are interfering destructively and the property of $MP_1$ interactions remains unaffected. 

For the region where magnon mode and second photon mode is getting coupled around 3.7~GHz, their respective internal and external dissipation and the coupling parameter between $M$ and $P_1$ i.e. $\Delta_{MP_1}$ are constant, so these parameters are not going to play much role in the transition of their behaviour. So the rest of the coupling parameters, namely $\Delta_{MP_2}$ and $\Delta_{P_1P_2}$ are dominantly and effectively playing the role in the transition phenomenon. 
For the Fig.~\ref{F3C5}(d, e, f) $\Delta_{MP_2}=0.1i$ and $\Delta_{P_1 P_2}=0$ so transmission profile, Fig.~\ref{F3C5}(d), and real part of eigenvalue, Fig.~\ref{F3C5}(e), both show CIA (LA), and their linewidth profile (imaginary part of complex eigenvalue), Fig.~\ref{F3C5}(f), shows repulsion.

For the Fig.~\ref{F3C5}(g, h, i) $\Delta_{MP_2}=0.01i$ and $\Delta_{P_1 P_2}=0.01i$ so transmission profile and real part of eigenvalue Fig.~\ref{F3C5}(g, h) respectively shows CIA while levels are just crossing, but their linewidth profile (imaginary part of complex eigenvalue) Fig.~\ref{F3C5}(i) shows very small reminisces of linewidth repulsion. So here for the Fig.~\ref{F3C5}(g, h, i) $\Delta_{MP_2}$ and $\Delta_{P_1 P_2}$ are dominating in such a way that level attraction approximately diminishes to level absorption which is clearly visible in Fig.~\ref{F3C5}(g, h) and also minutely visible in Fig.~\ref{F3C5}(i).
For the Fig.~\ref{F3C5}(j, k, l) $\Delta_{MP_2}=0.01i$ and $\Delta_{P_1 P_2}=0.1$ so the transmission profile and real part of eigenvalue, Fig.~\ref{F3C5}(j, k), respectively shows CIA behaviour while levels are still just crossing, but their linewidth profile, Fig.~\ref{F3C5}(l), totally diminishes repulsion behaviour and starts towards attracting the linewidth but still it is just about to cross. So here for the Fig.~\ref{F3C5}(j, k, l) $\Delta_{MP_2}$ and $\Delta_{P_1 P_2}$ are dominating in such a way that level attraction is diminished to level absorption which is clearly visible in Fig.~\ref{F3C5}(g, h) and also the signature of linewidth attraction which is just about to touch is visible in Fig.~\ref{F3C5}(i).
The outcomes of Fig.~\ref{F3C5}(g, h, i) and Fig.~\ref{F3C5}(j, k, l) signifies a special phenomenon known as level absorption which is visible in the midway when transition between level attraction  to repulsion is happening.

For the Fig.~\ref{F3C5}(m, n, o) $\Delta_{MP_2}=0.02i$ and $\Delta_{P_1 P_2}=0.2$. The transmission profile and real part of eigenvalue, Fig.~\ref{F3C5}(m, n), signifies CIT while levels are getting closer but remains splitted between higher and lower branches near the crossing center. Also in their linewidth profile, Fig.~\ref{F3C5}(o),  attraction in linewidth is clear as they are crossing each other. So here for the Fig.~\ref{F3C5}(m, n, o), $\Delta_{MP_2}$ and $\Delta_{P_1 P_2}$ are dominating in such a way that a clear coupling induced transparency appears in the Fig.~\ref{F3C5}(m, n) and also a clear  attraction in linewidth or crossing is visible in Fig.~\ref{F3C5}(o).

In the case of coherent coupling (LR/CIT) the frequencies of hybridised mode are repelling each other and their linewidths are crossing, but in the case of dissipative coupling (LA/CIA) the trends become opposite, i.e. frequencies of hybridised mode are attracting each other and their linewidths are repelling. But when system is undergoing transition from coherent to dissipative coupling or vice versa at the point when transition is happening the frequencies of hybridised modes are just crossing. Still it is characterised under coupling induced absorption but it is similar to the case when there is no coupling between the modes or there is very little coupling. So only by looking at the transmission and real part of eigenvalue profile it is hard to distinguish between absorption and no coupling. But even in this situation of CIA where levels are just crossing each other the existence of strong coupling near the coupling center signifies the phenomenon leading to blocking/ absorption of microwave transmission and also their linewidth profile branches still should be repulsive even for a narrow range of field. These two features together distinguish this transition point from weakly coupled or not coupled hybrid modes. Therefore in a multimode hybrid system the behaviour of the pair of modes under consideration not only depends on their parameters but may get influenced by the presence of other modes of the system too.

\section{Four modes coupled hybrid quantum systems}

\begin{figure}[h!]
    \centering
    \includegraphics[width = 0.85\linewidth]{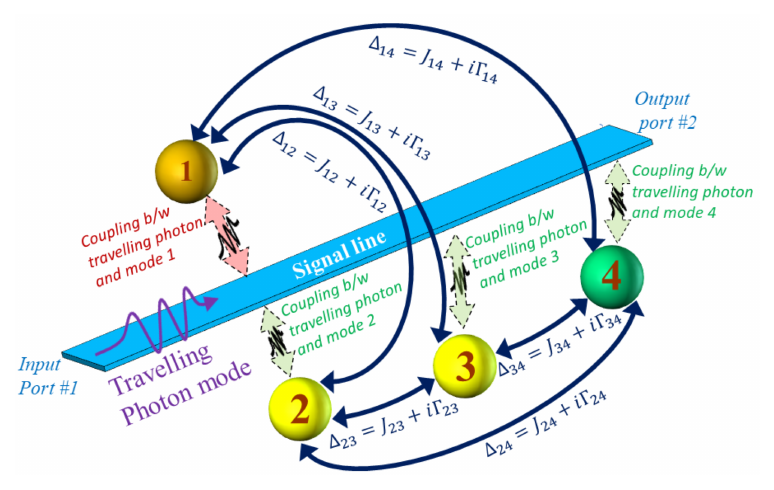}
    \caption
    {Cartoon showing a simple planar hybrid system having 4 modes, out of which mode~1 ($M$) is tunable and mode~2 ($P_1$), mode~3 ($P_2$) and mode~4 ($P_3$) are static modes. For tuning of $M$ we have applied an in-\,plane magnetic field perpendicular to the feedline/signal-line.}
    \label{F4C5}
    
\end{figure}

Motivated by a recent experiment \cite{bhoi2022coupling}, we now consider a relatively straightforward extension of the previous section with one more static mode. Here in total we have taken one tunable mode and three static modes in a planar hybrid structure. Schematic of the experiment is shown in Fig.~\ref{F4C5}. Similar to the previous section as tunable mode we are taking a magnon (YIG) that we are denoting by $M$, whose resonance frequency can be tuned by applying an external magnetic field. Static modes are photonic in nature and made of ISSRs, whose resonance frequencies are unaffected by the applied external magnetic field. For the three static modes we make three different ISRRs of different dimensions having different resonance frequencies. Like before the input and output ports of the VNA are attached to the MSL of the hybrid system through  which microwave photons are getting injected into the system resulting in all of the dynamics.

\subsection{Model and analysis}
After taking rotating wave approximation (RWA) we may write Eq.~\ref{E1C5} for four modes as~\cite{hu2022auxiliary, tiwari2024modified, rao2020interactions, harder2021coherent, walls2008quantum, scully1997quantum, shrivastava2024unveiling, manasi2018light}

\begin{align} \label{E15C5}
    H / \hbar &= \sum_{l=1}^{4}(\omega_l -i\alpha_l)\hat{X_l}^{\dag}\hat{X_l} +\sum_{1\leq l < m \leq N} \Delta_{lm}(\hat{X_l} \hat X_m^{\dag} + \hat X_m \hat X_l ^{\dag})  +\int\omega_k \hat p_k^\dag \hat p_k dk +\int\left[\sum\lambda_l (\hat{X_l} \hat p_k^\dag +\hat X_l^\dag \hat p_k)\right] dk 
\end{align} 
where different symbols have the same meaning as  the previous section with one change that here we have added a new photonic mode, fourth mode, in the system $P_3$ having resonance frequency $\omega_{P_3}$, corresponding creation (annihilation) operator being $\hat X_{P_3}^{\dag}(\hat X_{P_3})$, internal dissipation $\alpha_{P_3}$ and coupling parameters $\Delta_{MP_3}$, $\Delta_{P_1P_3}$ and $\Delta_{P_2P_3}$ for the coupling  between $MP_3$, $P_1P_3$ and $P_2P_3$ respectively. Also $\lambda_{P_3}$ is the strength at which travelling photons are driving the mode $P_3$. The external damping for the mode $P_3$ is $\beta_{P_3}$ giving effective damping for the mode $\alpha_{P_3}+\beta_{P_3}$. Following Eq.~\ref{E11C5} we can write $4\times 4$ coupling matrix as
\begin{align} \label{E16C5}
     \begin{bmatrix}
    &\tilde{\tilde{\omega}}_M 
    &\Delta_{MP_1}-i\sqrt{\beta_M \beta_{P_1}}
    &\Delta_{MP_2}-i\sqrt{\beta_M \beta_{P_2}}
    &\Delta_{MP_3}-i\sqrt{\beta_M \beta_{P_3}}\\
    &\Delta_{MP_1}-i\sqrt{\beta_M \beta_{P_1}}
    &\tilde{\tilde{\omega}}_{P_1}
    &\Delta_{P_1 P_2}-i\sqrt{\beta_{P_1} \beta_{P_2}}
    &\Delta_{P_1 P_3}-i\sqrt{\beta_{P_1} \beta_{P_3}}\\
    &\Delta_{MP_2}-i\sqrt{\beta_M \beta_{P_2}}
    &\Delta_{P_1 P_2}-i\sqrt{\beta_{P_1} \beta_{P_2}}
    &\tilde{\tilde{\omega}}_{P_2}
    &\Delta_{P_2 P_3}-i\sqrt{\beta_{P_2} \beta_{P_3}}\\
    &\Delta_{MP_3}-i\sqrt{\beta_M \beta_{P_3}}
    &\Delta_{P_1 P_3}-i\sqrt{\beta_{P_1} \beta_{P_3}}
    &\Delta_{P_2 P_3}-i\sqrt{\beta_{P_2} \beta_{P_3}}
    &\tilde{\tilde{\omega}}_{P_3}
     \end{bmatrix}_{4 \times 4} 
 \end{align}
where $\tilde{\tilde{\omega}}_{P_3}=\tilde{\omega}_{P_3} - i\beta_{P_3}=\omega_{P_3} - i(\alpha_{P_3} +\beta_{P_3})$. Complex eigenvalues of this matrix will give complex eigenfrequencies of this hybrid system.
In this case using  Eq.~\ref{E7C5}, for four modes we can write $P_{out}$ and $P_{in}$ relation in frequency domain as \cite{rao2020interactions,hu2022auxiliary,shrivastava2024unveiling}
\begin{align} \label{E12C5 }
\hat P_{out}(\omega)-\hat P_{in}(\omega)= - 2i \left[\sqrt{\beta_M} \hat{X}_M (\omega) + \sqrt{\beta_{P_1}} \hat X_{P_l} (\omega) + \sqrt{\beta_{P_2}} \hat X_{P_2} (\omega) + \sqrt{\beta_{P_3}} \hat X_{P_3} (\omega)\right] 
\end{align}
For dispersion spectra that takes into account the different types of cooperative effect with all the interactions of magnon mode, photon modes and travelling photon modes of the coupled hybrid system we need to calculate $S_{21}$ (on $\frac{\omega}{2\pi} - H_{dc}$ plane), which following Eq.~\ref{E8C5} can be written for 1 magnon and 3 photon modes as
\begin{align} \label{E17C5 }
S_{21}=\frac{\hat P_{out}}{\hat P_{in}}-1= - \frac{2i}{\hat P_{in}} \left[\sqrt{\beta_M} \hat{X}_M + \sqrt{\beta_{P_1}} \hat X_{P_1} + \sqrt{\beta_{P_2}} \hat X_{P_2} + \sqrt{\beta_{P_3}} \hat X_{P_3}\right]
\end{align}

\subsection{Results and discussion}

\begin{figure}
  \centering
  \includegraphics[width = 0.85\linewidth]{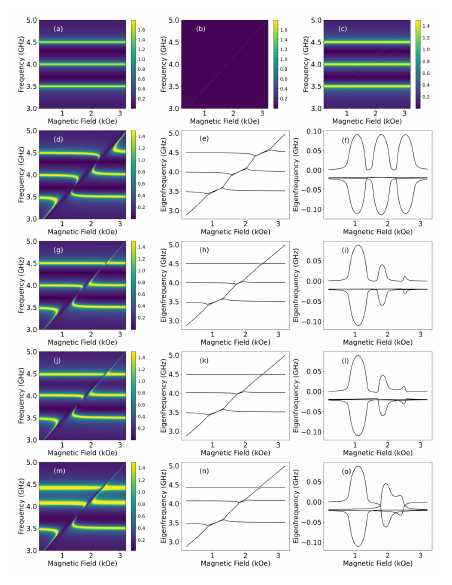}
    \caption{Hybrid quantum system having 1 TM coupled to 3 SMs.
    (a) Transmission profile for three static modes without magnon mode, (b) transmission profile for magnon mode without static modes, (c) combined transmission profile for three static modes and magnonic mode without any coupling.
   Transmission profile (d), (e) real, and (f) imaginary part of complex eigenfrequencies, when coupling between TM-2ndSM changed from $0.1i$ to $0.05i$, TM-3rdSM changed from $0.1i$ to $0.02i$.
   Transmission profile (g), (h) real, and (i) imaginary part of complex eigenfrequencies, when coupling between 2nd-3rdSMs changed from $0$ to $0.1i$.
   Transmission profile (j), (k) real, (l) imaginary part of complex eigenfrequencies, when coupling between 2nd-3rdSMs changed from $0.1$ to $0.2i$. The change of parameters and eigenvalue properties are summarized in Table~\ref{T2C5}.}
   \label{F5C5}
  \end{figure}

The results for this hybrid system are the expected generalisation of the previous case, but we have included it explicitly because this was the precise experimental setup of 
Bhoi et al.\cite{bhoi2022coupling}. We summarize the results in Table~2.

Fig.~\ref{F5C5}(a) shows the  $S_{21}$ power spectra of the three ISRRs without YIG film. Only three transmission lines horizontal to the applied $H_{dc}$ having resonance frequencies $\omega_{P_1}$= 3.5, $\omega_{P_2}$= 4.0 and $\omega_{P_3}$= 4.5 GHz with intrinsic damping rates $\alpha_{P_1}$= $\alpha_{P_2}$=$\alpha_{P_3}$=0.002 \cite{hu2022auxiliary,rao2020interactions,shrivastava2024unveiling} are visible in the transmission profile on the $(H_{dc} -f_{AC})$ plane which is unaffected by the applied magnetic field.  Fig.~\ref{F5C5}(b) showing $S_{21}$ power spectra of the YIG film without ISRRs, having intrinsic damping rate $\alpha_{M}$=0.00002 \cite{bhoi2022coupling, bernier2018level, hu2022auxiliary, rao2020interactions, harder2021coherent, shrivastava2024unveiling}, showing a slant transmission line which is getting tuned by the applied $H_{dc}$ ranging from $0.2$ to $3.2$ $kOe$. Resonance frequency of this FMR (magnon) mode can be linearly modelled as $\omega_{M} = 0.714\times H_{dc}+2.714$ which is varying from 2.856 to 4.998~GHz. Fig.~\ref{F5C5}(c) shows $S_{21}$ power spectra of the two ISRRs and YIG film but without any coupling, which is again a simple addition of Fig.~\ref{F5C5} (a) and (b) and transmission line of magnon is just crossing the transmission lines of photons.
\begin{table}[h!]
\begin{center}
\begin{tabular}{ |c|c|c|c|c|c| } 
 \hline
 Subplots & $\Delta_{MP_2}$ & $\Delta_{MP_3}$ & $\Delta_{P_2P_3}$ & \multicolumn{2}{c|}{Eigenvalues near $MP_3$ crossing}\\
 \cline{5-6}
 & & & & Real part & imaginary part \\
 \hline
 d-f & 0.1i & 0.1i & 0i & Attraction & Repulsion\\ 
 g-i & 0.05i & 0.02i & 0i & Intermediate & Repulsion\\ 
 j-l & 0.01i & 0.02i & 0.1i & Intermediate & Repulsion\\
 m-o & 0.02i & 0.2i & 0.2i & Repulsion & Attraction\\
 \hline
\end{tabular}
\end{center}
\caption{Set of parameters that are changing and the eigenvalues which are changing qualitatively for Fig.~\ref{F5C5}.}
    \label{T2C5}
\end{table}

Fig.~\ref{F5C5}(d, g, j, m) represents transmission profile, Fig.~\ref{F5C5}(e, h, k, n) and Fig.~\ref{F5C5}\,(f,~i,~l,~o) represent real 
and imaginary parts of complex eigenfrequencies respectively. 
In Fig.~\ref{F5C5} for each row (d, e, f), (g, h, i), (j, k, l) and (m, n, o) we can divide each sub-figure in three zones, first zone is for $MP_1$ around 3.5 GHz ($H_{dc}=1.1 KOe$), second zone is for $MP_2$ around 4.0 GHz ($H_{dc}=1.8 KOe$) and third zone is for $MP_3$ around 4.5 GHz ($H_{dc}=2.5 KOe$).
For Fig.~\ref{F5C5}(d, e, f), $\Delta_{MP_1} =\Delta_{MP_2}=\Delta_{MP_3}=0.1i$ and $\Delta_{P_1P_2} =\Delta_{P_1 P_3}=\Delta_{P_2 P_3}=0$ so for all three zones LA in observed in their transmission and real part of complex eigenvalue, and repulsion in their linewidth profile.

For Fig.~\ref{F5C5}(g, h, i) $\Delta_{P_1P_2} =\Delta_{P_1 P_3}=\Delta_{P_2 P_3}=0$, $\Delta_{MP_1}=0.1i$ is not changed although $\Delta_{MP_2}=0.05i$ and $\Delta_{MP_3}=0.02i$ is changed, their combined effect on zone 1 getting nullified so the nature of $MP_1$ interactions remains same. For zone 2 its LA nature in transmission and real part of complex eigenvalue profile have now diminished. Its reduced strength is also visible in their linewidth profile. For zone 3 its LA nature in transmission and real part of complex eigenvalue profile have now diminished to a greater degree and its much reduced strength is also visible in their linewidth profile.

For Fig.~\ref{F5C5}(j, k, l)  $\Delta_{MP_3} =\Delta_{P_1 P_3}=0$, $\Delta_{MP_1}=0.1i$, $\Delta_{MP_2}=0.05i$ and $\Delta_{MP_3}=0.02i$ is not changed and only $\Delta_{P_2 P_3}=0$ has changed to $0.1i$. Their combined effect on zone 1 again gets nullified, so the nature of $MP_1$ interaction remains the same. For zone 2 its LA nature in transmission and real part of the complex eigenvalue profile have now diminished a little more in strength that is also visible in their linewidth profile. For zone 3 the LA nature in transmission and real part of complex eigenvalue profile have now diminished to a greater degree. Its much reduced strength is also visible in their linewidth profile where repulsion is almost washed away and attraction or crossing is just about to start.

For Fig.~\ref{F5C5}(m, n, o)  $\Delta_{MP_3} =\Delta_{P_1 P_3}=0$, $\Delta_{MP_1}=0.1i$, $\Delta_{MP_2}=0.05i$ and $\Delta_{MP_3}=0.02i$ is again kept fixed only $\Delta_{P_2 P_3}=0$ has changed to $0.2i$. Their combined effect on zone 1 is again getting nullified so the nature of $MP_1$ interaction remains the same. For zone 2 the LA nature in transmission and real part of the complex eigenvalue profile have now little more diminished in strength that is also visible in their linewidth profile. For zone 3 the LA nature in transmission and real part of complex eigenvalue profile have ceased and LR has appeared. In their linewidth profile attraction also confirms CIT.

\section{Conclusion}

We have devised a comprehensive quantum theoretical framework for transition between CIA and CIT in a multimode coupled hybrid system which we have shown to work for three and four modes magnon-photon coupled interaction in a planar YIG-ISRRs coupled hybrid system. In a two mode coupled hybrid system the CIA and CIT are determined by dominant dissipative and coherent interactions respectively. Contrary to this in a multimode coupled hybrid system the local CIA\,/ CIT does not  only depend on the coupling parameter of the local hybrid modes involved but is a superposition of the all intra and inter coupling parameters of the system.
As a specific striking result of this, we are able to achieve CIT using only dissipative couplings, which in simpler case would have required a dominant coherent coupling.
This finding may be utilised to develop new insights for revealing origin of various effects, coupling interactions, and controllable transition between CIA to CIT and vice versa by manipulation of their coupling strength and dissipation rates etc. 
Although we have taken examples of magnon-photon coupled hybrid systems, our quantum formalism deals in terms of modes and these modes may be any other real or quasi-particles. This work may also open new experimental and theoretical pathways to explore similar phenomena on the other platforms that will help in advancement of such applications for various quantum devices and materials.

\section*{Acknowledgements}
The work was supported by the Council of Science and Technology, Uttar Pradesh (CSTUP), India, (Project Id: 2470, sanction No: CST/D-1520 and Project Id: 4482, sanction no: CST/D- 7/8). B Bhoi acknowledges support by the Science and Engineering Research Board (SERB) India- SRG/2023/001355

\section*{References}

%

\end{document}